# Detail reconstruction in binary ghost imaging by using point-by-point method

**Ning Zhang, Yanfeng Bai**\*, **Xuanpengfan Zou, and Xiquan Fu**

*College of Computer Science and Electronic Engineering, Hunan University, Changsha 410082, China*

*\*Corresponding author: yfbai@hnu.edu.cn*

**Abstract:** We propose a new local-binary ghost imaging by using point-by-point method. This method can compensate the degradation of imaging quality due to the loss of information during binarization process. The numerical and experimental results show that the target details can be reconstructed well by this method when compared with traditional ghost imaging. By comparing the differences of the speckle patterns from different binarization methods, we also give the corresponding explanation. Our results may have the potential applications in areas with high requirements for imaging details, such as target recognition.

# 1. Introduction

Ghost imaging [1-8] which reconstructs an unknown object by the correlation properties between two correlated beams, has attracted much attention because of its non-local property. In traditional ghost imaging (TGI) system, the object beam is usually collected by a bucket detector with no spatial resolution, and the reference beam is detected by a detector with spatial resolution, such as the charge-coupled device (CCD). The object can be reconstructed by measuring the intensity fluctuation correlation between the two detectors [9-14]. To improve imaging speed and imaging quality, ghost imaging technology has been widely investigated and combined with many new methods recently, such as compressive sensing GI [15-17], pseudo-inverse GI [18], normalized GI [19], spatial low-pass filtering [20-22], single-pixel imaging [23,24], deep-learning GI [25] and so on.

As we all know, TGI requires a lot of measurements to reconstruct high quality ghost-image. During this process, the acquirement, storage and transmission of large amount of data can affect imaging speed. For this problem, binarization ghost imaging (BGI) is put forward [26-29]. Chen et. al. indicated that grayscale object authentication based on ghost imaging using binary signals can possess a potential for advancing ghost imaging [26]. While, imaging quality by BGI is poor than that by TGI because the binary process is inevitably accompanied by the loss of information. To solve this problem, the group of Guo proposed a scheme in which imaging quality can be improved effectively by adding suitable random noise to the raw data before quantization [28], and our group demonstrated experimentally that Otsu binary ghost imaging (OBGI) can achieve a better imaging quality when compared with TGI [29]. In fact, the detail reconstruction for imaging object in the two works is not good. Note that edge detection based on ghost imaging which has been proposed recently, can present the detail edge information [30,31], but the low-frequency information is ignored.

In this paper, we present a new local-binary ghost imaging by point-by-point method (PPBGI). The basic idea of this method is to obtain the threshold corresponding to the current pixel point by comprehensively using the information of other pixel points, instead of simply using the fixed threshold to perform the binarization. It is shown that the details of the imaging object can be accurately reconstructed by setting the segmentation threshold corresponding to each pixel, which leads to the target information obtained by our method is richer than that from TGI. In addition, we also give a qualitative explanation of this phenomenon.

# 2. Model and theory

A TGI setup is shown in Fig. 1. The pseudo-thermal source is generated by illuminating a laser into a slowly rotating ground glass disk, then the light beam is divided by a beam splitter (BS) into a test and a reference beams, the test beam interacts with an unknown object, and be registered as an intensity sequence by a bucket detector. The intensity pattern of the reference beam which never touches the object, is recorded by a CCD camera synchronically. We can retrieve the object information by measuring the correlation function of intensity fluctuations between two detectors :

$$G(u_1, u_2) = \langle I_1(u_1) I_2(u_2) \rangle - \langle I_1(u_1) \rangle \langle I_2(u_2) \rangle, \tag{1}$$

where $u_i (i=1,2)$ is the transverse position of the $i$th detector, $\langle I_i(u_i) \rangle (i=1,2)$ donates the intensity distribution at the $i$th detector.

The information detected by the CCD camera in the reference path is crucial and sensitive to the reconstruction of the target object in ghost imaging, so we only quantify the data from the CCD camera. The key of binarization method lies in the selection of the threshold value. The earliest global binarization method which uses a unified standard to divide all pixels, includes the mean-binarization method and the maximum inter-class variance method (OTSU method) [32-35]. Here, OTSU method which calculates the optimal threshold (intra-class variance) through the characteristics of the histogram, has the advantages of simple algorithm, fast running speed and good effect, and become the most commonly used method in global binarization method. However, due to using a single threshold to partition all pixel points, the global binarization method makes the imaging information lost partly. Compared with traditional binarization method, point-by-point method sets the threshold for each pixel by combining global features and local features, which can express the image information more fully and reconstruct the image more accurately.

The threshold is set by the following method. The $m \times n$ pixel matrix $I$ of the reference beam is divided into several blocks with the size $k_1 \times k_2$: $I_{11}, I_{12}, \cdots, I_{1,\left[\frac{n}{k_2}\right]}, I_{21}, I_{22}, \cdots, I_{2,\left[\frac{n}{k_2}\right]}, \cdots\cdots, I_{\left[\frac{m}{k_1}\right],1}, I_{\left[\frac{m}{k_1}\right],2}, \cdots,$ $I_{\left[\frac{m}{k_1}\right]\left[\frac{n}{k_2}\right]}$. $T_{lh}(i,j)$ is set as the corresponding threshold of $I_{lh}(i,j)$, and $T_{lh}(1,1)$ is the initial value for further processing, and can be obtained by applying OTSU method to $I_{lh}$.

Assuming the distance between two adjacent pixel points as 1, the corresponding threshold of the pixel points in the first row and the first column can be calculated by using Eqs. (2) and (3).

$$T_{11}(1,j) = \frac{\frac{j-1}{k_2}}{1} \times T_{12}(1,1) + \frac{\frac{k_2-j+1}{k_2}}{1} \times T_{11}(1,1), \quad j=2,3,\cdots,k_2, \tag{2}$$

$$T_{11}(i,1) = \frac{\frac{i-1}{k_1}}{1} \times T_{21}(1,1) + \frac{\frac{k_1-i+1}{k_1}}{1} \times T_{11}(1,1), \quad i=2,3,\cdots,k_1. \tag{3}$$

Using the currently known threshold values, each row can be processed in turn through Eq. (4) to get the threshold corresponding to each pixel point in the block.

$$T_{11}(r,c) = \beta \times \sum_{i=1}^{r} \sum_{j=1}^{c} \left(1 - \frac{1}{(r+c)-1}\right)^{r-i+c-j} \times T_{11}(i,j), \quad r=3,4,\cdots,k_1; c=3,4,\cdots,k_2, \quad (4)$$

where $\beta$ depends on $k_1$ and $k_2$, and is used to normalize the coefficient in Eq. (4). So far, we have obtained the threshold values for all the pixel points in the first block. Performing the same process for all blocks, the threshold corresponding to each pixel points in the matrix can be obtained.

To prevent the boundary effect from the binary effect, we specify a global harmonic factor $\alpha$ to partition the original matrix by harmonizing the local threshold $T$ (obtained by the above steps) and the global threshold $t$ (obtained by applying OTSU method to the matrix $I$). Here, $\alpha$ depends on the requirement of details, and can be flexibly adjusted from 0 to 1 in practice. Binary segmentation of the original matrix is carried out:

$$I_{2'} = \begin{cases} 0, & I_2 \leq (1-\alpha)T + \alpha t \\ 1, & I_2 > (1-\alpha)T + \alpha t \end{cases}. \quad (5)$$

Now we get a new intensity distribution to recalculate Eq. (1):

$$G^{(2)}(u_1,u_2) = \langle I_1(u_1) I_{2'}(u_2) \rangle - \langle I_1(u_1) \rangle \langle I_{2'}(u_2) \rangle. \quad (6)$$

## 3. Results and discussion

To verify our method, the numerical simulation and the corresponding experiments are implemented in this section. During this process, the pseudo-thermal source was obtained by projecting a frequency-doubled pulsed Nd:YAG laser ($\lambda$=532nm and diameter $D$=6mm) onto a slowly rotating ground-glass disk. Firstly, we choose a simple double slit ($128 \times 128$ pixels, slit width $a = 0.2mm$ and separation $d = 0.6mm$) as the target object. With four different GI methods, Figure 2 presents the numerical (b1-d1) and experimental (b2-d2) results with 10000 measurements. By comparing Figs. 2(c) and (d) with (b), imaging quality by mean-binarization ghost imaging (MBGI) is poorer than that by TGI because of the information loss during the binary, while it can be improved by using OBGI [29]. Note that the reconstruction of the detail by the three methods is not good (see the part marked by the red circle). When PPBGI is applied, it is shown that the detail information is quite similar with the original object, as shown in Fig. 2(e). In addition, the experimental results agree well with the numerical patterns.

The detail information of the target object used in Fig. 2 is relatively simple. To highlight the advantages of our method, another more complex object, a pigeon with outstretched wings ($128 \times 128$ pixels) is further encoded by using the same parameters in Fig. 2, and the corresponding results are shown in Fig. 3. For TGI, MBGI and OBGI, the reconstructed results only present the rough information about the pigeon, and it is difficult to distinguish the feathers on the wings of the pigeon. However, the results are quite different when PPBGI is considered, our method can accurately recover the details, and one can clearly see the full outline of the pigeon, even the distinct feathers. By analyzing the results shown in Figs. 2 and 3, it is clear that PPBGI performs well in reconstructing the detail information, which may be helpful to present imaging details on the basis of reducing data size during the practical application of GI.

Next, the quantitative comparison between the results from four different methods is implemented by the correlation coefficient (Corr) between the target object ($O$) and reconstructed image ($G$)

$$\mathrm{Corr}(G,O) = \frac{\mathrm{Cov}(G,O)}{\mathrm{Var}(G)\mathrm{Var}(O)}, \qquad (7)$$

where Var($G$) and Var($O$) represent the variance, respectively. A large Corr value indicates better imaging quality. The Corr values of ghost-images presented in Figs. 2 and 3 are shown in Table 1. It can be seen that the Corr decreases by using the MBGI when compared with TGI. However, the Corr increases significantly after using OBGI and PPBGI, and imaging quality of PPBGI is always better than that from OBGI.

Finally, we give the corresponding explanation for the phenomena shown in the above. As we know, the resolution of ghost imaging is partly determined by the size of the speckle, and small speckle size corresponds to high resolution. Figure 4 shows the speckle pattern of the reference beam and the results after being processed by mean binarization, OTSU binarization and point-by-point binarization, and the corresponding experimental results of GI are also presented. Because of the small gap between the feathers of the pigeon, the high resolution is required to distinguish feathers clearly. In other words, we need the speckle with smaller size. From Fig. 4, it can be seen that the mean binarization makes some bright spots on the speckle pattern lose, which will lead to the loss of light field information and be not good for imaging process, as shown in Figs. 2(a2) and (b2). The speckle pattern is optimized by OTSU binarization (see Fig. 2(c1)), so imaging quality of OBGI is better [29]. Unlike the above methods which use the same threshold to divide all pixels, point-by-point method provides the corresponding threshold of each pixel, and makes full use of the threshold of the adjacent pixel points, which makes some speckle with smaller size occur (see the part marked by the red circle in Fig. 4(d1)). It is the reason why PPBGI can recover the target information more accurately and extract object details more fully. To verify our conclusion, we change the global harmonic factor $α$ from 0.15 in Fig.

4(d) to 0.4 in Fig. 4(e). It is clearly that some small speckles disappear by comparing Figs. 4(d1) and (e1), which results in the degradation of imaging resolution (see Figs. 4(d2) and (e2)).

## 4. Conclusions

In conclusion, we have proposed a new BGI scheme based on point-by-point method. It is proved numerically and experimentally that this method can improve the detail reconstruction of imaging target in ghost imaging, and provide more detail information than TGI. By comparing the changes of the speckle patterns after several binary methods, the corresponding reason is analyzed. Our method is more applicable for the complex targets imaged, and can be applied to some fields with the requirement of stricter imaging accuracy.

## Declaration of competing interest

The authors declare that they have no known competing financial interests or personal relationships that could have appeared to influence the work reported in this paper.

## Acknowledgements

The authors deeply acknowledge the support from the National Natural Science Foundation of China (NSFC) (Nos. 61871431 and 61971184), the Natural Science Foundation of Hunan Province (2017JJ1014). The authors express their gratitude to the editor and anonymous reviewers for the valuable comments and suggestions that helped us improving the manuscript.

# Figure captions

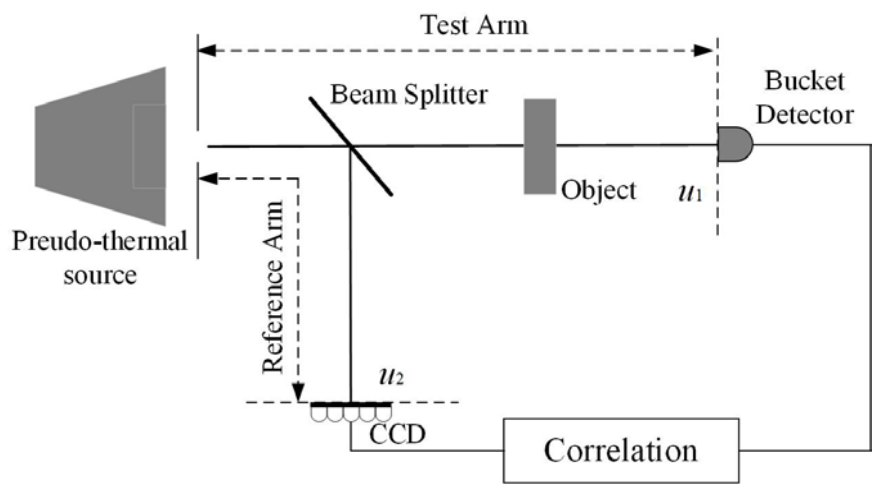

**Fig. 1.** A setup of ghost imaging with pseudo-thermal source.

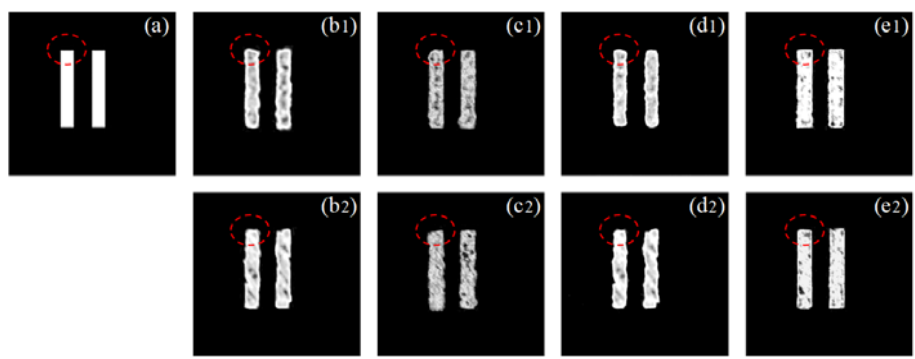

**Fig. 2.** A simple doubled-slit (a), and ghost-images by TGI (b), MBGI (c), OBGI (d), and PPBGI (e) with 10000 measurements. The first row presents the numerical results, and the experimental results are shown in the second row.

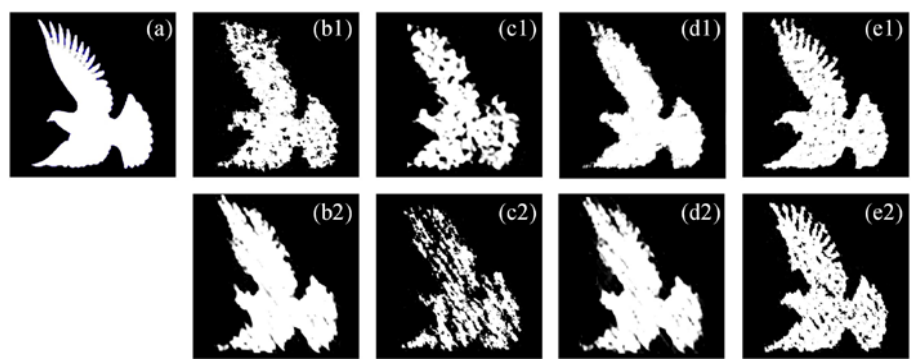

**Fig. 3.** The first row is the numerical results for a complex pigeon (a) by using TGI (b), MBGI (c), OBGI (d), and PPBGI (e) with 10000 measurements. The second row is the corresponding ghost-images in the experimental condition.

**Table 1. Results of correlation coefficient between object and ghost-image**

| Corr | Double slit | | Pigeon | |
|---|---|---|---|---|
| | Numerical result | Experimental result | Numerical result | Experimental result |
| TGI | 0.8639 | 0.8166 | 0.7929 | 0.7591 |
| MBGI | 0.8187 | 0.6701 | 0.7177 | 0.5096 |
| OBGI | 0.9032 | 0.8458 | 0.8243 | 0.7681 |
| PPBGI | 0.9241 | 0.8789 | 0.8931 | 0.8613 |

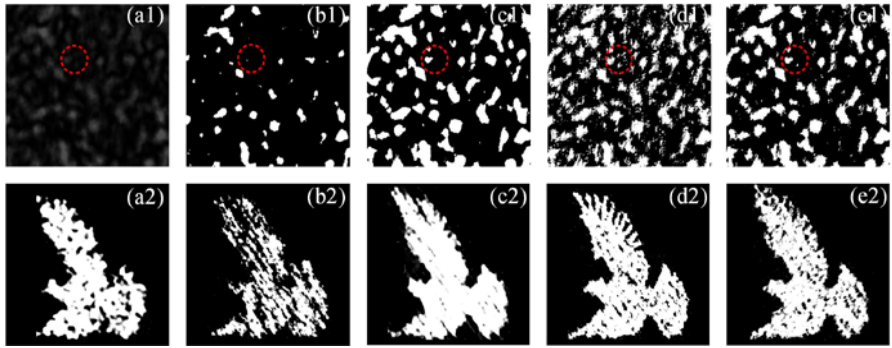

**Fig. 4.** The first row is the speckle pattern recorded on the reference detector (a1) and results processed by using MBGI (b1), OBGI (c1), PPBGI ($\alpha = 0.15$) (d1), and PPBGI ($\alpha = 0.4$) (e1). The second row is the corresponding experimental results.